# Practical Implementation of Memristor-Based Threshold Logic Gates

Georgios Papandroulidakis[1], Alexantrou Serb[1], Ali Khiat[1], Geoff V. Merrett[2], Themistoklis Prodromakis[1]

[1]Electronic Materials & Devices Research Group, Zepler Institute for Photonics and Nanoelectronics, University of Southampton, UK

[2]Cyber Physical Systems Research Group, School of Electronics and Computer Science, University of Southampton, UK

**Current advances in emerging memory technologies enable novel and unconventional computing architectures for high-performance and low-power electronic systems, capable of carrying out massively parallel operations at the edge. One emerging technology, ReRAM, also known to belong in the family of memristors (memory resistors), is gathering attention due to its attractive features for logic and in-memory computing; benefits which follow from its technological attributes, such as nanoscale dimensions, low power operation and multi-state programming. At the same time, design with CMOS is quickly reaching its physical and functional limitations, and further research towards novel logic families, such as Threshold Logic Gates (TLGs) is scoped. TLGs constitute a logic family known for its high-speed and low power consumption, yet rely on conventional transistor technology. Introducing memristors enables a more affordable reconfiguration capability of TLGs. Through this work, we are introducing a physical implementation of a memristor-based current-mode TLG (MCMTLG) circuit and validate its design and operation through multiple experimental setups. We demonstrate 2-input and 3-input MCMTLG configurations and showcase their reconfiguration capability. This is achieved by varying memristive weights arbitrarily for shaping the classification decision boundary, thus showing promise as an alternative hardware-friendly implementation of Artificial Neural Networks (ANNs). Through the employment of real memristor devices as the equivalent of synaptic weights in TLGs, we are realizing components that can be used towards an in-silico classifier.**

Today's conventional computing paradigm is based on the MOSFET transistor and CMOS technology; two cornerstones which have underpinned the development of digital electronics over the last 5 decades. Although there is still optimism for future improvement of CMOS, accumulating scientific evidence indicates the need for advances in both new emerging technologies to replace MOSFETs and in new computer circuits and architectures[1,2]. The former addresses the increasing difficulty of pursuing further downscaling (with its associated drop in reliability[2]) whilst the latter seeks to address the Von Neumann bottleneck, where increasingly big memories and powerful processors struggle to communicate over a limited interlink whose data transfer capacity doesn't scale fast enough[2–4].

On the computation/architecture front, there has been a sustained effort to develop bio-inspired computation concepts, mostly in the guise of artificial neural network-enabled (ANN) systems. Research on artificial neural networks has thus far spanned the entire interval between the first simplified models of all-or-none hardware neurons[5] and the current state-of-the-art GPU-based ANNs[6–8]. However, one often overlooked example of ANN-like computation can be found in the form of its quantized, digital counterpart, the so-called threshold logic (TL). TL is a model for performing a comparison between a threshold value and the weighted sum of an input vector. A basic computational unit in TL is called a threshold logic gate (TLG) and it corresponds to the artificial neuron in ANNs. Although TL is effectively a simplification of ANNs, TLG-based logic families have been shown to be capable of fast and low-power operation as evaluated by the power-delay trade-off metric[9–11]. Many of these implementations suggest the hardware implementation of ANN-like circuits and systems could greatly benefit from using TLGs as their fundamental logic cell, thus enabling advances towards neuromorphic computer architectures[12,13].

Many different designs of TLGs have been proposed with different trade-offs regarding their power-noise ratio performance[11] and different circuit implementations. Some introduce conventional electronics, such as capacitors and resistors, as part of their adaptive weights, while others take advantage of emerging nano-electronics devices such as single-electron technology (SET)[14] and negative resistance devices (NRD)[15,16]. However the majority of available TLG designs are based on transistor networks[11,17,18]. Recently, the use of Current-Mode (CM) Differential TLGs seems to gaining ground as one of the fastest and low-power differential threshold logic (TL) implementaions [11,19].

On the technology front, recent advances in emerging memory technologies introduce a new tool in electronic systems design. Specifically, ReRAM/memristor devices[20] act as nanoscale[21], finely tuneable[22], electrically programmable[23,24] resistive elements. Although, the transistor-based designs can provide some competitive solutions, their benefits are limited by increased sensitivity to noise and device mismatch, as well as limited fan-in (input vector space dimensionality) and relatively slow operation[25,26]. The memristors are capable of storing multi-bit information and retain their memory state even without power supply (non-volatile) while simultaneously their adoption in electronics is accelerated by additional advantages they provide such as better area scaling, low power consumption and CMOS-compatibility. Hence, memristor devices are considered as one of the most promising candidates for the next generation of computer circuits, systems and architectures[27–30]. This allows for area and power efficient reconfigurable circuit implementations, which are very important in a wide range of application areas, e.g. the embedded computing systems that process the data at the edge, where there is a continuous race towards minimization of chip area and power consumption for neuromorphic edge computing.

In this work, we demonstrate a practical implementation of the metal-oxide memristor-based CM-TLG (MCMTLG) circuits, thus laying the foundation for building artificial perceptron networks. We validate the functionality of the proposed gates through 2-input and 3-input experimental setups where tuneable memristor weights are used as artificial synaptic weights, defining the contribution of each input component to the TLG's comparison function. Notably, this functionality is enabled by the recently introduced multi-bit memristor technology[22], which enables fine, continuous control of the memristive synaptic weights. We show how changes in memristor resistances affect the shape of the decision boundary and comment on key, practical factors that affect performance.

**Fundamental components: TLG theory and memristor technology**

Initially, TLGs were introduced as a method of describing neural activity in the brain through conventional electronic circuits and systems[5]. Hence TLGs exist as a majority function logic technique able to perform linearly separable functions on a multi-dimensional, binary input space, a limitation induced based on the usage of digital electronic components. Just like a perceptron in single layer ANNs is able to separate a multi-dimensional continuous space into two parts divided by a line/hyperplane decision boundary, TLGs separate a binary input space using a line/hyperplane[31]. Notably, however, the decision boundary can be tuned more coarsely or finely regardless of the quantisation level of the input (or output) spaces, i.e. in a TLG, much like in an ANN, finely tuneable memristor resistance (memristance) values will lead to a finely tuneable decision boundary. Recent work moving towards optimized Current-Mode (CM) TLGs highlights the benefits this logic technology offers in comparison with conventional CMOS logic networks[10]. Furthermore, Dara et al.[19] has shown, through simulations, that a memristor-based concept can efficiently be incorporated into traditional differential TLG circuit design, hence becoming the catalyst of further power, noise sensitivity, logic and area scaling.

In this work we use a sort of memristive devices that has been shown to support analogue weights at a typical resolution of 5 bits (and max. res.: 7-bits)[32]. Prior to use in the memristor-enhanced CM-TLG circuits, the devices were pre-conditioned separately by electroforming and resistance stabilization in the required functional range. The devices present excellent retention characteristics with minimal state deviations even after long cycles of operation. Additionally to increased memory density, the multi-bit capabilities of the memristors can be used towards logic scaling[10,33] through the replacement of large MOS-based circuits with few nanoscale memristive devices, in technologies such as TLGs[34–37]. Furthermore, the use of memristors as continuously tuneable resistive devices is enabling the scaling in area and power of the TLGs, previously impracticable in VLSI technology due to limitation of the MOSFET devices. The flexibility and benefits of the memristive devices are responsible for introducing novel techniques towards enhancing the capabilities and performance of conventional TLG designs[38,39], with the memristors being an excellent nano-scale electronic counterpart for the synapses found in biological neural networks (NNs).

Combining an understanding of TLGs as fundamental computational units with the recently demonstrated multi-bit capabilities of memristive devices raises the prospect of a memristor-based reconfigurable fabric. In the following section we present initial experimental results using a discrete component-on-breadboard circuit implementation.

## RESULTS:

**Circuit design and operation**

The basis of our proposed hardware design, shown in Fig. 1a is based on the work of Dara et al.[19,33]. A CM-TLG design consists of two parts, the differential and the sensor part. The differential part consists of the input and threshold branches (Fig.1a), handling the input and threshold memristance input vectors respectively. Within each branch the weight vectors are implemented by a bank of 1T1M (memristively source-degenerated pMOS transistors) ensembles. Each 1T1M ensemble receives a digital input signal; a single element of the branch's input vector. If the input is low (active), then a memristance-dependent current flows from that input towards the sensor part. Additionally, each of the differential branches is power-gated by a back-to-back (BtB) pMOS circuit.

The sensor part is the thresholding element of the circuit, comparing the differential inflowing currents and settling to a binary output indicating which is greater. It is designed as an SRAM memory cell; a latching element consisting of two BtB connected CMOS inverters, forming a positive feedback loop (Fig.1a). Furthermore, two additional CMOS inverters are added at the outputs of the sensor part, one per output, hence avoiding any voltage level degradation and isolating the sensor part from the circuitry connected further down the logic cascade.

Overall, the MCMTLG circuit performs a current comparison operation in two phases. During the equalisation phase the differential part is power-gated on and the sensor part is power-gated off, thus forcing the sensor part into an unstable equilibrium. In that phase the voltages at CO and CA are forced to be almost equal by the shunting BtB pMOS devices between the branches. Next, in the evaluation phase the inter-branch shunting is released, the differential part is power-gated off and and the sensor part is power-gated on. This allows differences on nodes CO and CA to be amplified by the positive feedback action of the BtB-connected inverters of the sensor part[25]. Notably the differential part is cut-off from the voltage supply during the evaluation phase, thus disabling the current flow towards the sensor, leaving only a brief window for the sensor of achieving a stable and correct transition to a binary memory state, based on the small voltage differences settled during the equalization.

**Circuit validation**

The MCMTLG was built using discrete components on a breadboard. Similar to Dual Clocked Current Mode Threshold Logic Gate (DCCML) design[33] we used a common voltage supply for both sensor and differential parts. Furthermore, the differential 1T1M banks were connected to the outputs of the sensor, thus speeding up the sensor decision-making operation regarding the performed current comparison by removing the RC paths introduced by the differential path. This practice is similar to the coupled inverters with asymmetrical loads (CIAL) technique[40]. For our physical implementation with discrete components, the differential circuit uses two pMOS transistors back-to-back for power-gating. This is done to control the connection of the input and threshold vectors to the reading voltage supply, thus avoiding logic state degradation of the latching element during evaluation phase, as well as improved operational stability even with noisy input vectors. The BtB pMOS circuits also enable lower power consumption, due to the fact that the differential part is cutoff and does not consume power during the evaluation phase. The voltage supply $V_{DD}$ used in the proposed design is 0.65V, thus ensuring that the memristive devices being use cannot be accidentally programmed during TLG operation. Furthermore, a BtB PMOS circuit was used also for the equalization circuit that reset the sensor part before the evaluation being performed. $V_{CLK}$, which control the operation cycle of equalization/evaluation, and the input vector's high voltage levels, are set to 0.9V. The low logic level for both the $V_{CLK}$ and the input vector is set to 0V (Gnd). A microcontroller, (Raspberry Pi 3 Model B) is used to generate the clock signal as well as the input vector used for the experiments described below.

In the 2-input circuit experiments (measured results in Fig.1d-h), the threshold branch consists of a single 1T1M sub-circuit (TH) while the input vector consists of two 1T1M sub-circuits (M1, M2). SImultaneously, AND and OR functionality, can be interpreted as different flavours of majority-gating. For example a MAJ-1 gate is equivalent to a 2-input OR gate and a MAJ-2 is homomorphic to an AND gate. This implies memristive weights set in such way that either input has a larger weight than the threshold branch (i.e. M1,M2 < TH). Similarly for AND: M1,M2>TH, but M1||M2<TH. This can be extended to much more complex threshold functions [19,37].

For the 2-input AND TLG case we used the weight configuration of {M1, M2; TH} = {60.5kΩ, 60kΩ; 33kΩ}, thus satisfying the requirements for the AND TL inequality equation (majority-2 function). From the canonical (CA) output we can measure the AND function circuit response, while from the complementary (CO) output we can obtain the complementary function, NAND. The measured response of the AND/NAND TLG configuration is shown in Fig.1c and the input vector is presented in

Fig.1e for the first input (IN1) and in Fig.1f for the second input (IN2). The clock signal that controls operation is shown in Fig.1g. Due to the use of binary input vectors the quantized corner points of the 2D area (1b,c)are of interest. The CA (canonical) output, where AND and OR functions can be measured, and CO (complementary) output, where the NAND and NOR functions can be measured, can be seen in Fig.1a. Fig 1c showcase the complementary (CO) output of the circuit configured to perform the 2-input AND/NAND gate (NAND(CO)), while in Fig.1d the measured CO output for the OR/NOR configuration (NOR(CO)) is shown. The clock signal is determining the equalization/reset phase (clock HIGH) and evaluation/set phase (clock LOW) cycle of operation. The outputs NAND(CO) and NOR(CO) are valid during the evaluation phase, while during the equalization we can see that the output signals stay at an intermediate unstable logic level. The $V_{DD}$=0.65V and the $V_{CLK}$=$V_{IN}$=0.9V (for the logic '1'). It is worth noting that due to the use of pMOS devices in the 1T1M sub-circuits of the differential part, the logic for HIGH input voltage the corresponding input is non-conductive (logic '0') while for LOW input voltage the corresponding input is conductive (logic '1').

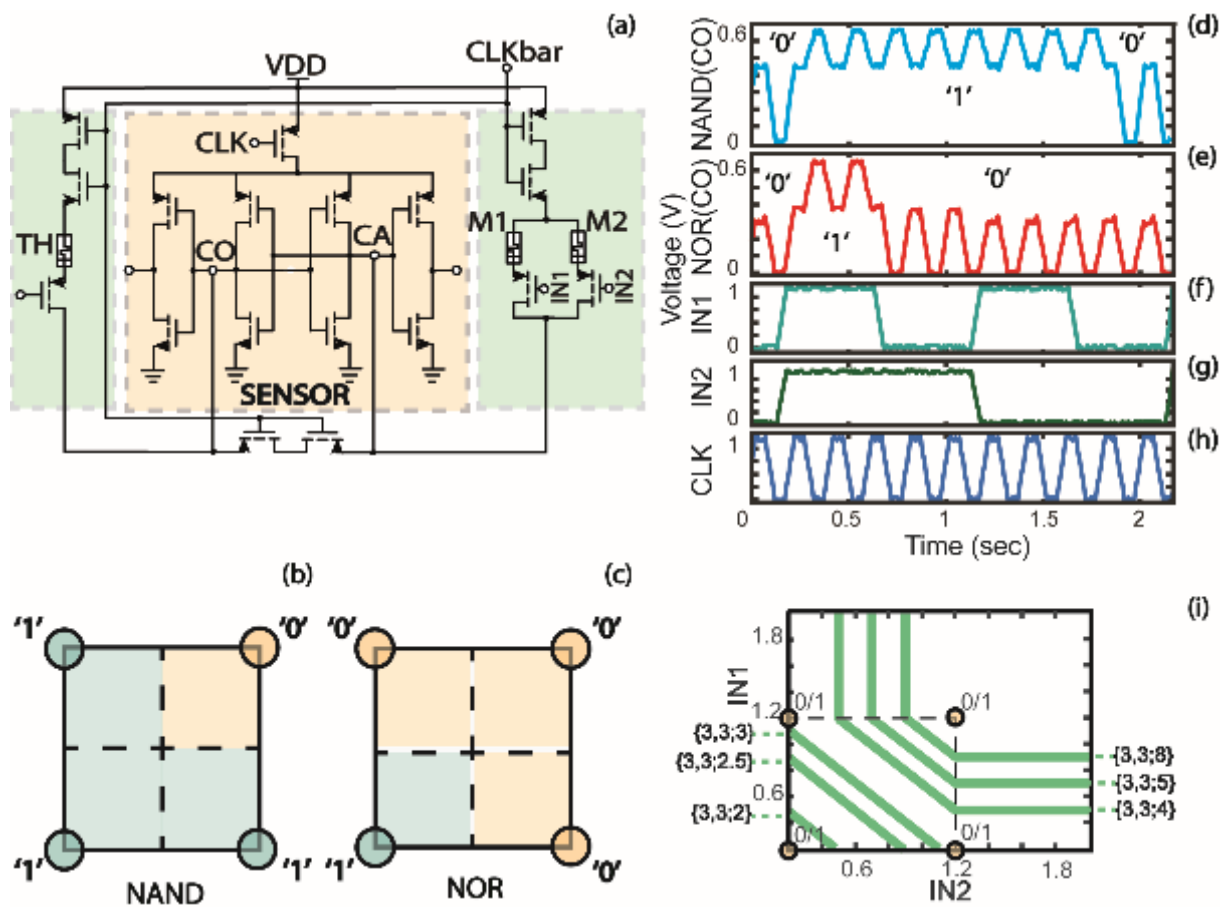

**Figure 1: Memristive Current Mode Threshold Logic Gate (MCMTLG) 2-input configurations case study**, Circuit schematic; **a**, indicative 2D area splitting of quantized space performed by the 2-input MCMTLG, for the AND case; **b**, and the OR case; **c**, experimentally measured results from the hardware implemented MCMTLG for the evaluation of the AND/NAND, where the AND complementary output (NAND(CO)) is showcased as measured from the Complementary (CO) output of the sensor; **d**, and OR/NOR function, where the OR complementary output (NOR(CO)) is showcased as measured from the Complementary (CO) output of the sensor; **e**, control signals; **f, g, h,** simulation of a 2-input MCMTLG emulator showing shifting of the 2D input plane decision boundary due to changes in the threshold weight; **i**. The different configuration performed by the MCMTLG were enabled by the memristor configurations of {M1, M2; TH} = {*60.5kΩ, 60kΩ; 33kΩ*} {*33kΩ, 18.3kΩ; 41.7kΩ*} for the AND/NAND and OR/NOR function, respectively. Regarding the Fig.1i, each weight configurations denoted on the side of the each case, in the form of {a, b: c}, is weight multipliers where the unit is in MΩ.

In our case the differential part configuration was set as {M1, M2; TH} = {33.8kΩ, 18.3kΩ; 41.6kΩ} for the OR function. These values are chosen from the available dynamic range of the memristor programmability. Similarly to the configuration of the AND/NAND case study, the input vector (Fig.1e-f) and the clock signal (Fig. 1g) are the same for this operational configuration. Both OR and NOR, as well as the AND/NAND, functions are performed simultaneously due to the complementary bi-stable operation of the sensor part.

To highlight the reconfigurability of the MCMTLG and its effect on the defined classification boundary, additional SPICE simulations (LTSpice simulation environment) were performed using an emulator version of the proposed circuit. In Fig.1i different plane classification thresholds are showcased by changing the threshold memristive weigth values of the 2-input MCMTLG emulator configuration and recording the decision boundary for each case. In the case of an digital sensor part, like the one used in the proposed design, the first three memristor configurations ({3MΩ, 3MΩ; 2MΩ}, {3MΩ, 3MΩ; 2.5MΩ}, {3MΩ, 3MΩ; 3MΩ}) will result in an OR/NOR logic gate while the remaining three configurations ({3MΩ, 3MΩ; 4MΩ}, {3MΩ, 3MΩ; 5MΩ}, {3MΩ, 3MΩ; 8MΩ}) in an AND/NAND logic gate. The different memristance scale used in the simulations (in MΩ), compared to the hardware experiemental setups (in kΩ), is based on our goal of testing the MCMTLG for the MΩ memristance range. The high current limiting effect of these memristances in parallel arrays enables the implementation of process invariant TL circuits[41].

One of the most important advantages that the TLGs computing paradigms offer is the computing of complex Boolean functions with virtually the same circuit just by assuming the use of larger parallel 1T1M networks. Fig.2a shows, a 3-input MCMTLG design, where a third 1T1M sub-circuit has been added to the input branch of the differential part and memristors are employed as the analogue weights at the input/bias binary signals, similar to the 2-input MCMTLG. Introducing the memristor devices as analogue weights in a digital logic gate family, has the advantage of enabling highly localised, continuously tuneable, minimal front-end footprint and low-voltage operated non-volatile memory into the TLG, thus providing a potentially decisive advantage in the implementation of memory-heavy ANN accelerators [37].

The measured results (Fig.2c-d) are extracted from a 3-element input vector and 1-element threshold setting of MCMTLG. The upper trace of Fig.2c shows the response for {M1, M2, M3; TH} = {31.5kΩ, 30kΩ, 28.2kΩ; 68.2kΩ} memristive weight configuration which functions as a 3-input OR, where at least one of the input sub-circuit need to be conductive in order for the current comparison to result in an input-side winning node (majority-1 function). In Fig.2c the measured response for the canonical output (CA) while in Fig.2d the output of the isolation inverter connected to the CA output is shown. We can observe that during the equalization phase the voltage level or the isolation inverter is 0V while performing a full voltage swing to logic '1' when the corresponding CA output drops to logic '0'. The input vector and clock signal are shown in Fig. 2e-h. In the case of memristor devices use in this configuration, similarly to the 2-input case described above, the same circuit could perform multiple functions just by reprogramming the threshold resistive value, thus configure differently the winning conditions of the current comparison performed.

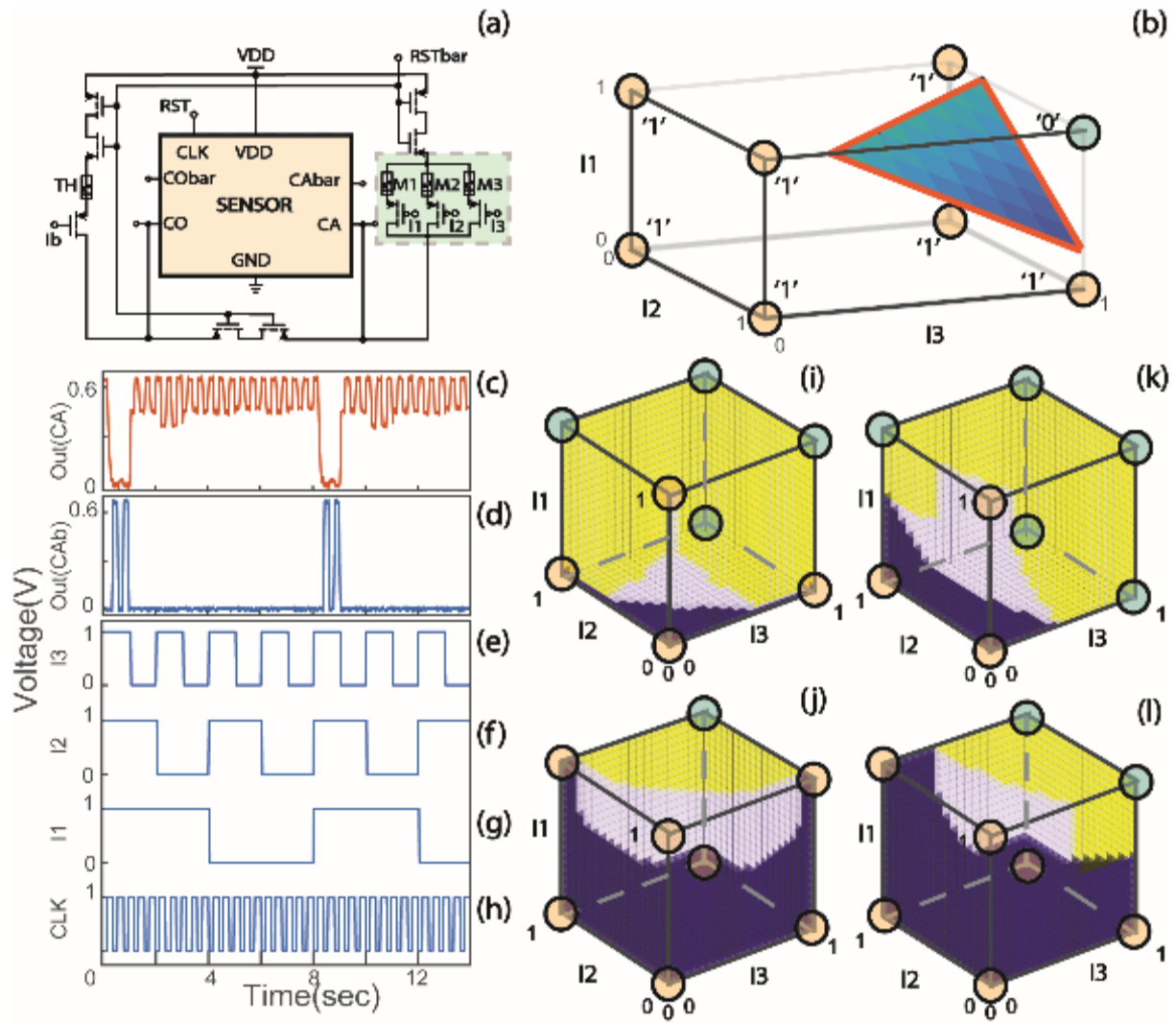

**Figure 2: Practical implementation of 3-input MCMTLG experiment.** Schematic of a 3-input MCMTLG, where the sensor component includes the CMOS latching element and the two additional restoration inverters per output; **a**, 3D plane splitting of space performed by the circuit; **b**, canonical (CA) output of the circuit configuration; **c**, output of the isolation inverter feeded the CA signal (CAbar); **d**, the control signals (clock and 3-input vector {I1, I2, I3}); **e, f, g, h,** 3D plane splitting of a simulated MCMTLG emulator with symmetric input weights; **I, j**, 3d plane splitting of a simulated MCMTLG emulator with asymmetric input weights; **k, l**. The weight configuration for the OR3/NOR3 is {M1, M2, M3; TH}={31.5kΩ, 30kΩ, 28.2kΩ; 68.2kΩ}.The 3D quantized space separation is shown next to the circuit schematic (2b), where the corner cases of the 3D space are evaluated through the operation of the MCMTLG.The measured results for the OR3/NOR3 configuration (3c) are from the canonical (CA) output of the sensor part. In (3d) the ouput of the isolation inverted, feeded from the CA output, CAbar is showcased. For the input vector HIGH is logic '0' while LOW is logic '1', due to the use of pMOS devices in the 1T1M. Similarly to Fig. 1i, a SPICE MCMTLG emulator was used to evaluate the changes in the 3D plane splitting behavior of the circuit for symmetric input weights and assymetric input weights. Here the weights configuration used to obtain the simulated results were: {2MΩ, 2MΩ, 2MΩ; 1MΩ }, for the fig. 2i,{2MΩ, 2MΩ, 2MΩ; 2.5MΩ }, for the fig. 2j, {8MΩ, 2MΩ, 4MΩ; 2MΩ }, for the fig. 2k and { 8MΩ, 2MΩ, 4MΩ; 4MΩ }, for the fig. 2l.

Similarly to the operation of a perceptron or other simple ANN we can observe that the memristive synaptic weights are responsible for the plane splitting of the 2D or 3D space. Through arbitrary reconfiguration of the memristive values we are able to shift the decision boundary of neural decision-making functionality and thus alter the system of inequalities that the MCMTLG solves. Taking all the above into account, we could conclude that MCMTLGs are ideal for a higher level memory-centric reconfigurable fabric implementation, where the functionality of a computational fabric is controlled by the ReRAM memory contents distributed into this sea of gates. Thus, to further strengthen the concept of reconfigurable computing, enabled by this circuit design, a series of SPICE simulations, in

LTSpice environment, have been performed using an emulator version of the physically implemented circuit where for different values of the memristive weights we are able to separate the plane with different classification areas, as shown in Fig. 2i-l. Two cases are explored and for each case the input weights are kept constant. For the first case the input weights configuration is set to a symmetric configuration of {2MΩ, 2MΩ, 2MΩ} while the threshold weights is changed from 1MΩ to 2.5MΩ. Similarly, for the second case the input weight is set to an asymmetric configuration of {8MΩ, 2MΩ, 4MΩ} while the threshold weights is changed from 2MΩ to 4MΩ.

## CONCLUSIONS:

In this work we have presented a physical implementation of a proof-of-concept TLG circuit using reconfigurable memristive loads enabling mixed signal weight elements implementation of the threshold and inputs networks. Through the presented experimental results we have shown that the comparison operation between the memristance of the threshold device and the composite impedance of the input network defines the circuit functionality. One of the most important drawbacks regarding the adaption of TL-based circuit designs into VLSI systems and computer architectures is the lack of sufficient tools to synthesize TLGs into a modern system design[9,42]. More improvements in this area need to be proposed, and the hardware realization of such circuit will provide a clear understanding regarding the methods we need to apply for their application. Advances towards novel memristive TLG systems design have the potential to provide benefits to low-optimized digital circuitry such as CMOS implementations of popular operations like convolution, encryption and other bitwise-based operations.

The initial results of the proposed proof of concept circuits and systems indicate that the technology being studied can potentially have great impact on the physical implementation of artificial neural networks inside SoC designs through truly mixed signal implementation, hence a departure from the conventional Convolutional NNs. Thus one of the most prominent application of this logic technology is the implementation of unconventional more-than-Moore computer circuits, systems and architectures such as novel neuromorphic computing systems. Following this work, we should focus on scaling up the input vector and testing the fan-in capabilities of the proposed design. At the same time, it is of great importance to exploit the benefits of different memristive technologies and capture the full picture of what each memristor devices can provide by being incorporated into a MCMTLG circuit. This future work will provide important tools towards developing primitive TL computing blocks for hardware realization of more complex systems.

## Methods

**Memristive device fabrication and specification**: All the memristors used in the experimental setups are in 3×3mm$^2$ chips that are wire-bonded to PLCC68 packages. Each memristor is a 20x20 um$^2$ cross-point of top and bottom electrodes. All circuits implemented throughout this work rely on the rich dynamics of an in-house metal-oxide ReRAM technology employing metal-insulator-metal (MIM) devices. Originally, the devices were fabricated on 6-inch SiO$_2$/Si wafer with bottom and top electrodes (BEs & TEs) patterned using optical lithography, e-beam evaporation and liftoff processes. Similar processes were adopted for the active layer patterning, except that sputtering was used for the deposition with a magnetron-sputtering tool. The active layer is constituted of TiO$_2$ and Al$_2$O$_3$ thin-film metal-oxides. After dicing, 3x3 mm$^2$ wire-bonded chips containing memristor devices were obtained, with MIM stacks constituting of Pt/Al$_2$O$_3$/TiO$_2$/Pt/Ti (10/4/24/10/5) nm. Fig. 2.a shows an example of a chip that contains 32 stand-alone devices, with dimensions of 60x60 μm$^2$. Scalability is another advantage of memristors when used in cross-bar array configuration. Pi et al. have demonstrated devices down to 2x2 nm$^2$ with 12 nm pitch[43]. Nonetheless, scalability of resistive switching devices using arrays is reaching the full potential only after resolving sneak-current issue by introducing appropriate selector devices.

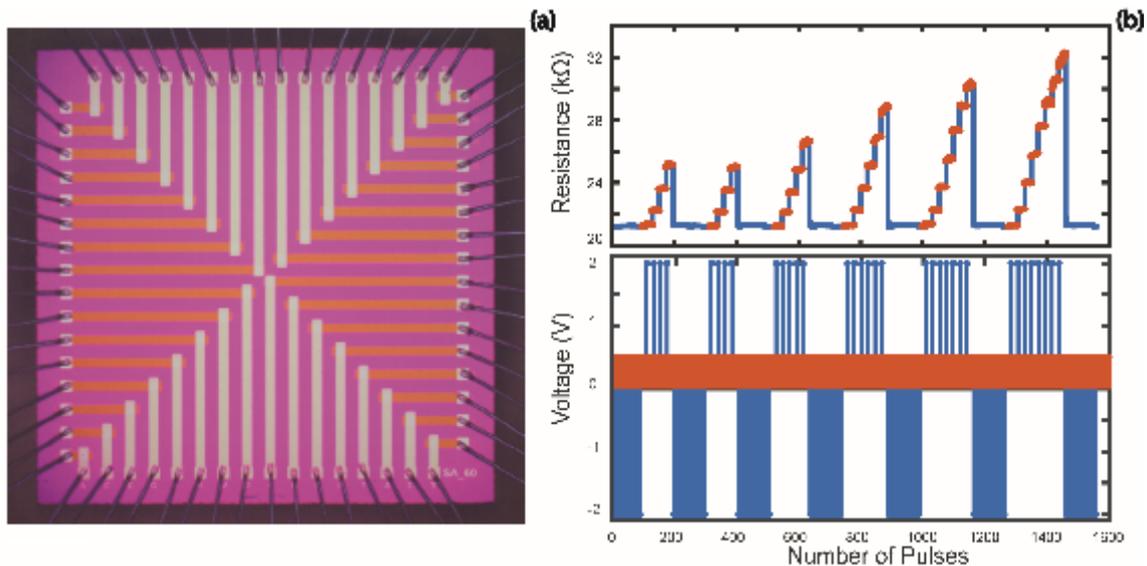

**Figure 3: Characteristic behavior of the memristor devices.** a standalone memristive topology is shown, where each crosspoint device is a memristor; **a**, arbitrary programming of select memristive states [8]; **b**. The upper trace of (**b**) presents the memristance programming due to the voltage bias pulse, shown in the lower trace of (**b**). The memristors are fabricated in-house from our team Alongside the +2V bias pulses corresponding to the multi-state memristance programming, we can observe the +0.5V reading pulses, showcased in orange color, as well as the -2V pulses that assure us that before each programming phase the memristance is resetted fully.

**Experimental set-ups and procedures**: All hardware experiments illustrates in Fig.1 and Fig.2 were carried out on circuits prototyped on breadboard. External power supply was used to supply the power rail of the implemented circuits. The results were gathered exclusively by oscilloscope. For these experiments packaged devices were used, connected to the breadboard discrete component-based circuit using a breakout board. The power supply used for the experiments was 0.65V, to avoid any unwanted state programming through the trains of reading pulses applied to the differential part of the circuit. For the pMOS devices we used the NDP5020P (1H10AA) model while for the nMOS devices we used the SUP85N02-03 (T32BAA) model. The memristor devices used throughout these experiments are detailed above. We measured the hybrid circuit response through the Rigol MSO4000 the response of the practically implemented memristor-based TLG. The input vector and the clock signal were produced through microcontroller programming and converter to circuit-specific voltage levels through custom resistive voltage divider circuits. It worth noting that for the case of 3-input we measured the input vectors and the clock signal through the Logic Analyser (LA) digital probes, due to the limited number of analogue probes available from the oscilloscope. In each experiment, the memristive devices used were programmed in the required state using an ArC ONE instrument board (ArC Instruments, UK). All devices used for all the experiments were located on the same die, i.e. only one memristive device package containing a total of 32 memristors.

**Acknowledgements:** This work has been supported by the Engineering and Physical Sciences Research Council (EPSRC) grants EP/K017829/1.

**Author contributions:** GP, AS and TP conceived the study, GP ran the experiments, processed the resulting data, AK developed the process flow and fabricated the memristive devices. All authors contributed to writing and refining the manuscript.

**Competing financial interests:** The authors declare no conflict of interest of any kind.

**Data availability:** All relevant data is available from the authors. Furthermore this is deposited at TBD with DOI TBD.